# Impact of $Al_2O_3$ Passivation on the Photovoltaic Performance of Vertical $WSe_2$ Schottky Junction Solar Cells

*Elaine McVay [‡, \*], Ahmad Zubair [‡], Yuxuan Lin, Amirhasan Nourbakhsh and Tomás Palacios*

Department of Electrical Engineering Computer Science, Massachusetts Institute of Technology, 77 Massachusetts Avenue, Cambridge, Massachusetts 02139, United States

.

ABSTRACT

Transition metal dichalcogenide (TMD) materials have emerged as promising candidates for thin film solar cells due to their wide bandgap range across the visible wavelengths, high absorption coefficient and ease of integration with both arbitrary substrates as well as conventional semiconductor technologies. However, reported TMD-based solar cells suffer from relatively low external quantum efficiencies (EQE) and low open circuit voltage due to unoptimized design and device fabrication. This paper studies Pt/$WSe_2$ vertical Schottky junction solar cells with various $WSe_2$ thicknesses in order to find the optimum absorber thickness. Also, we show that the devices' photovoltaic performance can be improved via $Al_2O_3$ passivation which increases the EQE by up to 29.5 % at 410 nm wavelength incident light. The overall resulting short circuit current improves through antireflection coating, surface doping, and surface trap passivation effects. Thanks to the $Al_2O_3$ coating, this work demonstrates a device with open circuit voltage ($V_{OC}$) of 380 mV and short circuit current density ($J_{SC}$) of 10.7 mA/cm$^2$. Finally, the impact of Schottky barrier height inhomogeneity at the Pt/WSe$_2$ contact is investigated as a source of open circuit voltage lowering in these devices.

**Main Text**

Transition metal dichalcogenide (TMD) materials are emerging candidates for flexible thin-film photovoltaics to be used in applications such as internet of things, flexible electronics, and ubiquitous electronics[1] due to their high absorption coefficient[2-4], mechanical flexibility[5], high carrier mobility and ultrathin nature[6]. Thanks to their weak van der Waals interlayer bonding, these materials (in their ideal form) offer a self-passivated surface free of dangling bonds that can also be heterogeneously integrated with bulk materials as tandem solar cells to maximize their photovoltaic efficiency.

Tungsten diselenide (WSe$_2$) is one of the most promising TMD materials due to its bulk bandgap of ~1.3 eV[7] that offers maximum photovoltaic efficiency for a single absorber according to the Shockley-Queisser formalism[8]. The high electron and hole carrier mobility (greater than 100 cm$^2$/Vs)[9] and high absorption coefficient (i.e. $10^5$ cm$^{-1}$ at 780 nm[10] compared to silicon's $10^3$ cm$^{-1}$) make it an attractive option for high performance photovoltaic applications. Most of the earlier research efforts on TMD photovoltaics focus on the monolayer limit due the direct nature of the bandgap. However, monolayer TMDs can only absorb ~10% of the incident light[6] and, although this absorption could be increased with appropriate resonators, it is arguably more straightforward to simply use a thicker absorber layer. Recent works show that multilayer WSe$_2$ and other TMD materials can offer promising performance for thin-film photovoltaic applications. The ideal approach for TMD solar cell design is to use p-n junction diodes[11-12]. However, due to the lack of a reliable substitutional doping technology in TMD materials, research efforts have been mostly focused on Schottky junction solar cells. These devices offer a relatively simple fabrication process and the option to tune the open circuit voltage by work function engineering of the contact metal. Jariwala *et al.*[3] demonstrated that excellent (higher than 90%) absorption characteristics can be achieved for multilayer WSe$_2$ flakes (~15 nm) by optimized photonic design (silver back reflector to maximize the absorption via

Fabry-Perot effects). Recently, Went *et. al.* demonstrated a novel metallization technique to improve the interface between the metal and the TMD absorber layer. The authors show that transferred gold onto 16 nm $WS_2$ can form a large Schottky barrier leading to a PCE of 0.46 % and open circuit voltage of 0.256 V at AM1.5 illumination. Si *et al.* studied several contacts metals to $WSe_2$ and showed that the high Schottky barrier between Zn contacts and ~100-nm-thick $WSe_2$ can increase the open circuit voltage (~ 0.35 V), but did not explore the effects of absorber layer thickness.

Despite the excellent progress in optimizing the absorption and photonic design in $WSe_2$ Schottky junction solar cells, the demonstrated solar cell power conversion efficiency (0.8-6.7% measured at 532nm illumination)[3, 13] is still well below the expected theoretical values (20% – 27%)[2]. To realize the ultimate potential of layered transition metal dichalcogenides in thin film photovoltaics, it is necessary to also carefully design and understand the electronic transport, doping and passivation technologies in these devices[2, 14-15].

This work demonstrates high work function $Pt/WSe_2$ vertical Schottky junction solar cell diodes using multi-layer $WSe_2$ as the absorber layer. We explore the trade-off between absorber thickness and device performance as a function of observed $V_{oc}$ and $J_{sc}$ to identify the optimal thickness range (80-150 nm) for Schottky junction solar cell applications. We also study the impact of $Al_2O_3$ passivation on the fabricated $WSe_2$ solar cell performance. The passivated devices show dramatic improvement of the external quantum efficiency and short circuit current without impacting the open circuit voltage. Finally, we investigate the Schottky barrier height distribution of our Platinum-to-$WSe_2$ contact that acts to lower the effective Schottky barrier height, thus decreasing the open circuit voltage due to an increase in dark current. Although further in-depth study is required to understand the effect of passivation and how to increase the uniformity of the Schottky contacts, our findings reveal a promising direction for improving the performance of the layered TMDC solar cells.

A schematic of the device used in our study is shown in Figure 1a. The work function difference between the bottom electrode and semiconductor absorber layer determines the open circuit voltage of the devices. The theoretical large work function difference (0.5 eV) between platinum and n-type $WSe_2$ makes this metal an attractive option for the bottom electrode. Mechanically exfoliated and dry-transferred $WSe_2$ serves as the absorber layer with patterned gold electrode on top for carrier collection. The gold electrodes form a nearly ohmic contact due to their negligible Schottky barrier height with $WSe_2$ (See supplementary Figure S2 for Au/$WSe_2$/Au *I-V* characteristics.)[16]. A layer of electron-beam patterned hydrogen silsesquioxane (HSQ) serves as the isolation between top contact and sidewalls of $WSe_2$ to prevent in-plane conduction through the $WSe_2$ which may contribute to the dark current. The device was then capped with an aluminum oxide ($Al_2O_3$) layer for passivation and charge transfer doping studies. The $Al_2O_3$ is grown via a two-step deposition process utilizing an electron-beam-evaporated seed layer and subsequent atomic layer deposition (ALD) layer. A list of specific devices discussed in the main text are summarized in Table 1, and detailed processing information is included in the Experimental Methods section.

Table 1: Summary of devices discussed in the main text

| Device | Thickness of $WSe_2$ absorber layer ($t_{WSe2}$) |
|--------|--------------------------------------------------|
| A      | 180 nm                                           |
| B      | 120 nm                                           |
| C      | 74 nm                                            |
| D      | 25 nm                                            |

Figure 1b shows the dark current-voltage characteristics of device A, a representative $WSe_2$ Schottky diode with $t_{WSe2} \approx 180$ nm. The diode exhibits a well-defined rectifying behavior in the dark with a rectification ratio (current at 1 V of applied bias compared to -1 V applied bias) greater than $10^4$. Under AM1.5 illumination, the device shows photovoltaic performance with open circuit voltage ($V_{OC}$) of 380 mV, short circuit current density ($J_{SC}$) of 10.7 mA/cm$^2$, fill factor of 44% and maximum generated power of 1.5 mW/cm$^2$ (see supplementary Figure S3, supplementary Figure S4, and supplementary Table S1). Figure 1c compares the simulated absorption of WSe$_2$ flakes of various thicknesses as calculated by the transfer matrix method[17] using n, k data extracted from the literature[18] (see details in supplementary section 5). It demonstrates a high broadband absorption, even down to the 10 nm thick flake, of WSe$_2$ on platinum. However, the experimental results of fabricated devices shown in Figure 1d demonstrate better short-circuit current performance in the devices with a total thickness in the 80 nm-150 nm range. The relatively lower $J_{sc}$ for absorber layers thinner than 120 nm can be attributed to imperfect separation of photogenerated carriers produced near the metal/semiconductor interface. Similar $J_{sc}$ lowering trends due to insufficient carrier separation were observed in published simulations that studied the thickness scaling of Si solar cells[19]. Meanwhile, the $J_{sc}$ rolloff for flakes thicker than 120 nm comes from a change in the absorption characteristics. We have included a more detailed discussion on the $J_{sc}$ and $V_{oc}$ and a function of WSe$_2$ thickness in supplementary section 6 with reference to supplementary Figure S5. We use devices B and C that are within the optimal intermediate thickness range for detailed passivation studies.

In conventional and thin film solar cells, passivation[14-15] can further improve device performance and stability. Passivation plays an important role in semiconductor devices by protecting the surface, removing defects, and managing the electric field profile, as well as acting as an anti-reflection coating. Recent works have shown that Al$_2$O$_3$ can be used to passivate and n-type dope MoS$_2$ transistors[16] and photodetectors[20]. Al$_2$O$_3$ was also used to improve the PCE of a MoS$_2$/Si solar cell, although the dominant mechanism for the improved passivation was not investigated[21]. In addition to these devices, a related work uses the hole transport layer WO$_x$ on WSe$_2$ to induce p-type doping in a heterojunction solar cell[22]. Here we systematically study the impact of Al$_2$O$_3$ passivation on Schottky junction WSe$_2$ solar cell performance. We identify and quantify three mechanisms through which Al$_2$O$_3$ passivation improves device performance: surface trap passivation, surface doping, and antireflection coating effect.

The effect of the Al$_2$O$_3$ passivation layer thickness on the solar cell performance was investigated in detail on device B with the optimum WSe$_2$ thickness ($t_{WSe2}$= 120 nm). Figure 2a shows the change in the dark current due to Al$_2$O$_3$. The high forward bias region shows little change after passivation, (series resistance is unchanged), and the ideality factor remains at ~1.28 (see supplementary Table S2) for all passivation thicknesses, suggesting that bulk recombination processes are unchanged in the diode. However, the low forward bias (highlighted in Figure S6a) and reverse bias regions (highlighted in Figure S6b) show suppression of dark current upon passivation. Figure 2b shows the change in photovoltaic characteristics as a function of the Al$_2$O$_3$ passivation layer thickness. The initial 2 nm electron-beam-evaporated Al$_2$O$_3$ shows the largest relative improvement (69 %) in $J_{SC}$ from 18.3 nA to 31 nA. The further addition of Al$_2$O$_3$ passivation shows relatively smaller improvement in $J_{SC}$. An additional 5 nm ALD Al$_2$O$_3$ shows 8.3 % increment in $J_{SC}$, while the relative improvement is slightly larger (11.3 %) for another 10 nm of ALD Al$_2$O$_3$ compared with the 5 nm ALD Al$_2$O$_3$. This suppression of the reverse bias dark current along with the increase in short circuit current can explain the improvement in the open circuit voltage seen in Figure 2b from 0.22 V to 0.25 V. Further, the largest relative improvement due to passivation is seen between the unpassivated case and the 2 nm AlO$_x$ cases, and additional AlO$_x$ thickness provides incremental improvement to the short circuit current.

From the results of Figure 2b the short circuit current could be improved due to any combination of passivation of trap states, doping of the tungsten diselenide, and anti-reflection coating effect of the Al$_2$O$_3$. To identify the origin of short circuit current improvement, the absorbance characteristics of the WSe$_2$ flake is measured under broadband white light source before passivation and compared with 2 nm electron-beam-evaporated Al$_2$O$_3$, and additional 20 nm ALD passivation. As shown in Figure 2c and supplementary Figure S7a, the change in the absorbance due to the 2 nm passivation is negligible (~1%) while the improvement due to the additional 20 nm ALD passivation reaches a value of 12% at the 550 nm wavelength. The simulated absorbance in Figure 2d and supplementary Figure S7b for different passivation thicknesses confirms the experimental observations. The marginal improvement in the absorption due to 2 nm AlO$_x$ compared to the 69.4% increase in the short circuit current indicates that the anti-reflection coating effect of the $Al_2O_3$ cannot alone explain the large increase in short circuit current. Instead, we conclude from this experiment that the 2 nm $AlO_x$ coating improves the device performance through a combination of surface doping and surface trap passivation.

To further understand the effect of $Al_2O_3$ on the short circuit current improvement, we map photocurrent under short circuit conditions ($V_a$=0 V) and measure the external quantum efficiency (EQE) at the 532 nm wavelength for device C ($t_{WSe2}$=74 nm). Figure 3a shows the top-view optical image of the device. A cross section of the normalized photocurrent (symbols) along the A'-A line depicted in Figure 3b shows relatively higher and more uniform photocurrent collection away from the top electrodes (ohmic contacts) in the passivated device. To study the effect of surface doping, the device was modeled using Silvaco Atlas (See supplementary section 9 and Table S5 for details). Interestingly, the simulated curve for $2\times10^{11}$ cm$^{-2}$ surface doping shows a good match with our experimental results (black symbols), suggesting that unpassivated devices may have some fixed positive charge at the surface. The simulated curves also show that increases in positive surface charge lead to large improvements in photocurrent collection away from the top electrode. The improvement in photocurrent monotonically increases with the positive surface charge density. The simulation further indicates that the passivated case can be qualitatively matched with ~$2\times10^{12}$ cm$^{-2}$ surface doping, suggesting that the $AlO_x$ does indeed add charge to dope the surface of the $WSe_2$. This value is in agreement with the literature, which suggests that $AlO_x$ can be used to dope surfaces with up to $4\times10^{12}$ cm$^{-2}$ positive charge[23]. Likewise, the normalized photocurrent maps for pristine and passivated devices presented in Figures 3c and 3d, respectively, show an enlargement of the effective device active area for photocurrent generation after the $Al_2O_3$ passivation. As the electric field across an absorber layer is increased, the collected photocurrent as a function of lateral position becomes more uniform and the decay profile broadens due to an increase in the drift length $L_E = \mu E \tau$, where, $\tau$, $\mu$ and $E$ are the lifetime, mobility and the electric field.[24-25] Figure S9 shows that the electric field increases monotonically through the flake along the BB' line, with a 24% increase in the electric field at 40 nm from the surface when comparing the $2\times10^{11}$ cm$^{-2}$ and $2\times10^{12}$ cm$^{-2}$ doping cases. The resultant increase in $L_E$ allows for more efficient minority carrier separation. Figure 3e shows the cross section potential profile of the passivated device simulated using Silvaco Atlas to further study the impact of the surface doping on the electrostatics. The positive surface doping induced by the $Al_2O_3$ contributes additional n-type band bending (and electric field) near the surface of the device, as shown in Fig 3f. The electric field engineering near the surface of the device due to the n-type charge transfer doping increases the area available for photocurrent extraction for this device geometry. Interestingly, supplementary Figure S10 shows that for input powers of 0.1 µW/µm$^2$ and lower, the $1\times10^{12}$ cm$^{-2}$ positive surface charge doping can be used to yield 95% or greater normalized photocurrent across the WSe$_2$ absorber over a distance of 18 µm away from the Au electrode. The results of Figure 3 and supplementary Figure S10 suggest that the $Al_2O_3$ passivation could allow for more widely spaced gold electrodes in an optimized design.

Figures 4a and 4b demonstrate the wavelength dependence of the device's absorption and EQE with and without passivation. Both absorption and EQE measurements show peaks in the data near 760 nm due to the A-exciton[26] in bulk WSe$_2$. The B-exciton is located between 550 nm and 600 nm[27], which likewise contributes to the increase seen in the EQE and absorption characteristics[10] in this wavelength range. We note that the broadband improvement in the absorption in device C is not the same as the improvement in the absorption in device B due to different Fabry-Perot interference patterns for material stacks with different thicknesses. However, the absorption improvement is again consistent with simulation as shown in supplementary Figure S11. Because both devices were fabricated according to the same process and use WSe$_2$ flakes extracted from the same crystal, we expect the surface doping and surface trap passivation effects to remain the same. Figure 4c shows the relative change in the EQE compared to the relative change in the absorption. Here, it can be seen that between the 650 nm and 700 nm wavelengths the EQE improvement is mostly due to the $Al_2O_3$'s antireflection coating effect. At higher wavelengths and near the 760 nm A exciton peak we observe large enhancement in the EQE that is not attributed simply to absorption improvement. From the discussion on Figure 3, we can point to surface doping as one mechanism to improve the EQE. However, Figure S12 also shows through simulation that the photocurrent magnitude near the Au cathode will increase if the surface recombination (srv) is decreased. Therefore, we suggest that this wavelength dependent enlargement of the EQE is due to the surface doping identified by the photocurrent spatial mapping as well as the passivation of traps that reduce the surface recombination.

Both the passivated and pristine EQE results shown in Figure 4b highlight the need to improve the EQE at longer wavelengths to achieve high efficiency devices. The EQE reaches a maximum of 29.5% at 410 nm, but is below 20% at 750 nm, which still is a wavelength above the indirect bandgap of ~1.31 eV (947 nm). We additionally observed an EQE of 38.4% at 410 nm and 19.7% at 750 nm at a different measurement location from the one tracked in Figure 4 (see Supplementary Figure S12). A tandem solar cell with a material that more efficiently absorbs at the longer wavelengths or incorporates plasmonic structures, up-converting materials, or passivation layers that improve the device performance in this long wavelength region can potentially overcome this limitation.

In an ideal Schottky junction solar cell, the maximum expected open circuit voltage is approximately equal to the Schottky barrier height between the metal and the semiconductor. Although this work reports a high Schottky diode $V_{oc}$ value at AM1.5 illumination (up to 0.38 V), this experimentally observed value is small considering platinum's work function of ~ 6.0 eV[28] and $WSe_2$'s electron affinity of ~ 4.0 eV[7]. To investigate the origin of the discrepancy between expected and experimentally observed $V_{OC}$, we perform systematic temperature dependent transport measurement of the $WSe_2$ Schottky diode. Due to the van der Waals nature of the interlayer bonding, the vertical series resistance of the 180 nm thick $WSe_2$ diode is expected to be very high and it may lead to inaccurate extraction of the Schottky diode parameters. To overcome this limitation, we fabricated a device on a thinner $WSe_2$ layer, device D ($t_{WSe2}$ =25 nm). The analysis was repeated in supplementary Figure S14 for a device based on a 150 nm flake (device E) which shows a similar trend, but the result is affected by the large $R_S$. Neither device D nor device E were passivated as we do not expect that $Al_2O_3$ passivation on the top surface will affect the Pt/$WSe_2$/Au junction.

Figure 5a shows the temperature dependent transport characteristics of device D. Temperature dependence of the reverse bias characteristics suggest the existence of trap-assisted tunneling processes[29] due to a low Schottky barrier height (see Supplementary Figure S15), while the temperature dependence of the forward bias region indicates thermionic emission dominated current. From the Arrhenius plot[30], a maximum Schottky barrier height of 0.23 eV is extracted for Pt/WSe$_2$. This barrier height is lower compared to the previously demonstrated experimental values and suggests the existence of Fermi pinning as well as barrier height inhomogeneities[31]. The low barrier height limits our maximum obtainable V$_{oc}$ and has also been observed in transferred Au/WSe$_2$ contacts[32]. Contrary to the Pt/WSe$_2$ case, the ebeam lithography defined Au/WSe$_2$ contact is not dominated by thermionic emission as can be seen from its weak temperature dependence (see supplementary table S7), allowing us to apply the single Schottky junction model in our analysis.

Schottky barrier inhomogeneity is a common phenomenon in bulk semiconductor devices[31, 33] but is typically ignored in the analysis of layered materials devices. The barrier between metal and a semiconductor can be inhomogeneous due to the fabrication process or polycrystalline nature of the semiconductor. Ignoring Schottky barrier inhomogeneity may lead to inaccurate extraction of the barrier height. The level of inhomogeneity can be extracted from the ideality factor at low and high biases by using the methodology of Werner and Güttler[31].

The ideality factor was extracted from both the low bias region, (via $dI/dV$) and the high bias region (see supplementary table S7) (via Norde plot), demonstrating that the ideality factor decreases with temperature and has a value of 1.45 at room temperature as shown in Figure 5b. The ideality factor of a Schottky diode with inhomogeneous Schottky barrier can be expressed as,

$$\frac{1}{n(T)} - 1 = \frac{\rho_2}{\frac{2k_bT}{q}} - \rho_1 \qquad (1)$$

Where n(T) is the temperature dependent ideality factor, $\rho_1$ is the difference between the mean barrier height at zero bias and the mean barrier height at applied bias, $V_a$ and $\rho_2$ is the difference between the standard deviation at zero bias and the standard deviation at applied bias, $V_a$. Figure 5c shows that $\frac{1}{n(T)} - 1$ is always positive, similar to PtSi/Si Schottky diodes,[31] indicating the effective barrier height increases with increasing bias. We extract the values of $\rho_1$ and $\rho_2$ coefficients to be .0896 and -0.0111, respectively from the linear fitting. The negative value of $\rho_2$ indicates that barrier height distribution homogenizes with increasing bias. In comparison to Werner's PtSi/Si Schottky diodes, our Pt/WSe$_2$ diodes show a 2× higher $\rho_2$ coefficient, suggesting that this diodes' barrier height distribution is strongly affected by the applied bias. This observation indicates that the Schottky barrier height inhomogeneities lower the effective barrier height within the regime of solar cell operation, thus lowering the open circuit voltage from its potentially higher value.

Figure 5d shows the barrier height increases monotonically with the applied bias even after the barrier height estimation is corrected for inhomogeneities. A secondary conduction mechanism, i.e. parallel leakage path[34], causes the effective Schottky barrier height to decrease with decreasing voltage. At high bias, the thermionic emission current dominates, increasing the Schottky barrier height until it eventually saturates (Figure 5d). This leakage path contributes to additional open circuit voltage lowering.

**Conclusion**

In this work, we demonstrate high work function Pt/WSe$_2$ vertical Schottky junction solar cell diodes using multi-layer WSe$_2$ as the absorber layer. We identify the 80-150 nm WSe$_2$ thickness range as optimal for photovoltaic performance, investigate the effect of Al$_2$O$_3$ passivation on device performance, and evaluate the nature of Schottky barrier height inhomogeneities that reduce the effective barrier height and solar cell $V_{oc}$. The effects of Al$_2$O$_3$ passivation on solar cell characteristics shows clear short circuit improvement due to an improved EQE and photocurrent collection area. The improvement in EQE can be attributed to passivation of trap states, anti-reflection coating effects, and an altered carrier collection pathway due to doping. We show 50% higher EQE at wavelengths below 600 nm. More broadly, surface passivation techniques such as the one presented here can be used as a facile way to tune the performance of a TMD optoelectronic device for a desired application.

**Experimental methods**

Schottky diodes were fabricated on silicon wafers with 90 nm $SiO_2$ thermal oxide on top for convenient flake identification. Titanium (5 nm)/Platinum (45 nm) electrodes were patterned via electron beam lithography and electron beam deposition. The surface of the platinum layer was cleaned with Nanostrip®, 50:1 $HF:H_2O$, and ozone for 2 minutes prior to flake transfer. $WSe_2$ flakes were extracted from a bulk crystal from HQ Graphene by micromechanical exfoliation and subsequently transferred on top of the pre-patterned electrodes using a pickup and dry transfer method[35]. After transfer, the flakes were cleaned in hot acetone followed by 350°C annealing in $N_2$ environment for 3.5 hours to remove polymer residue from the dry transfer process. This step is necessary to ensure higher quality top contact. Next, an isolation border was patterned onto the edge of the $WSe_2$ flake using electron beam exposed Hydrogen Silsesquioxane (HSQ) spin coated onto the sample to prevent conduction through the edge of the flake which may have different properties than the bulk vertical conduction[36]. Then, ohmic contacts were patterned onto the $WSe_2$ flakes using electron beam lithography. Gold contacts were deposited in low pressure ($<5 \times 10^{-7}$ torr) to reduce the contact resistance of the device[16]. Finally, $Al_2O_3$ coatings were applied to all devices except device D. The first 2 nm of passivation is deposited as aluminum via electron beam evaporation to act as a seeding layer for subsequent ALD layers to promote growth uniformity. The electrical characterization was performed using a B1500 semiconductor parameter analyzer. The low temperature measurements were conducted in a Lakeshore cryogenic probe station using liquid nitrogen as a coolant. Solar simulator measurements were conducted at an AM1.5 work station with a calibrated lamp.

For the scanning photocurrent measurements, a broadband supercontinuum fiber laser (Fianium) was combined with a monochromator to generate the monochromic laser beam with the desired wavelength. A two-axis piezo-controlled scanning mirror was coupled to a microscope objective through two confocal lenses to perform the spatial scanning with the laser beam spot of around 1 μm on the device. The photocurrent and the reflected light intensity were recorded simultaneously to form the scanning photocurrent images and the reflectance images. The photocurrent results were measured under the short-circuit condition, in which zero voltage bias was applied across the device. The absorbance and EQE measurements were carried on with the same setup. The incident laser power was measured at the output of the microscope objective using a calibrated photodetector, and the reflected laser power was measured with another photodetector at a divided beam path location after a beam splitter that was inserted before the input of the objective. The reflectance was calculated by normalizing the measured reflected power spectrum on the devices by the measured reflected power spectrum on a silver mirror.

FIGURES

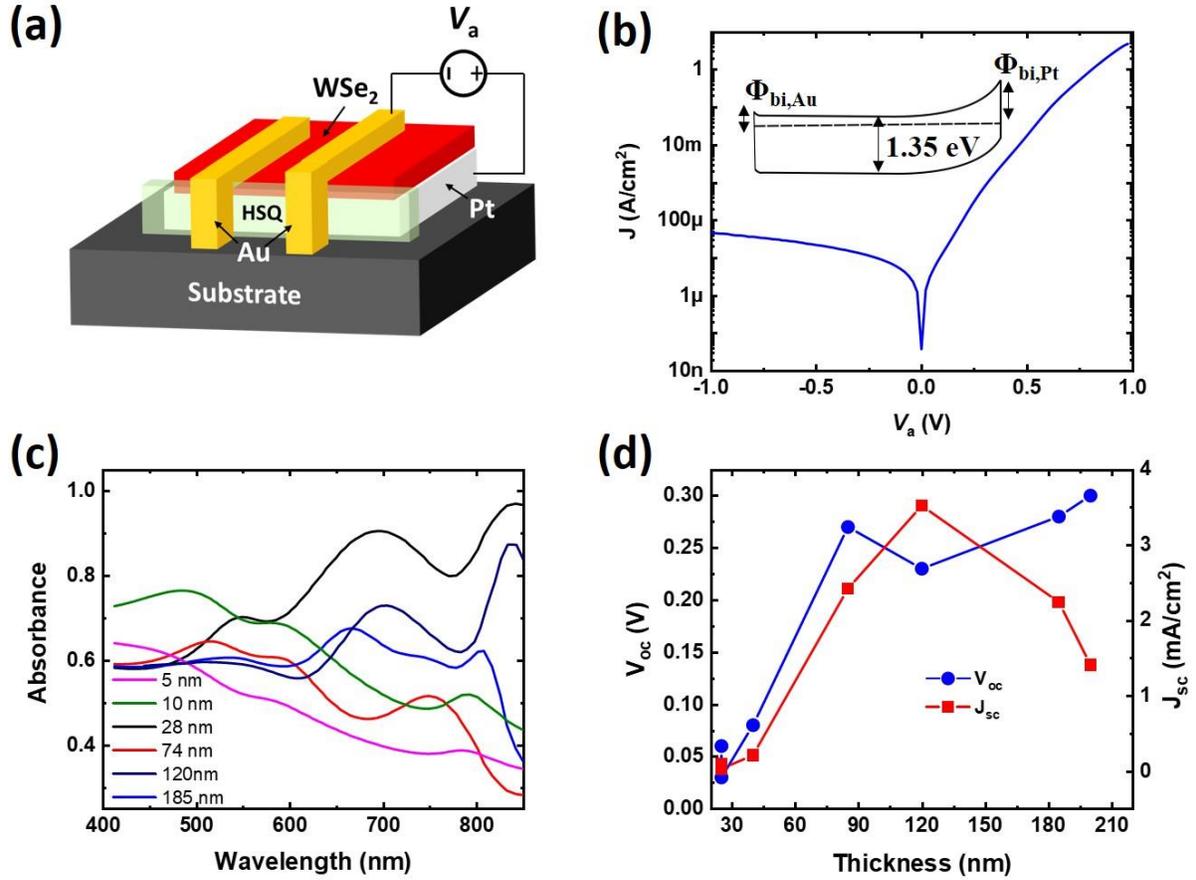

Figure 1. (a) Schematic diagram of the fabricated Schottky barrier solar cell, where a $WSe_2$ layer is sandwiched between top (Au) and bottom (Pt) metal contacts. An electron beam cross-linked HSQ is used as isolation dielectric between the two contacts. (b) Transport characteristics of a typical $WSe_2$ diode (device A, 180 nm) show a rectification ratio>$10^4$ at 1 V compared to -1 V. Inset shows the equilibrium ($V_a$=0 V) energy band diagram under illumination. (c) Simulated absorption characteristics of layered $WSe_2$ on Pt back contact for different film thickness. (d) Open circuit voltage and short circuit current as a function of $WSe_2$ thickness (including devices A-D in the main text) under 0.3 $W/m^2$ broadband illumination.

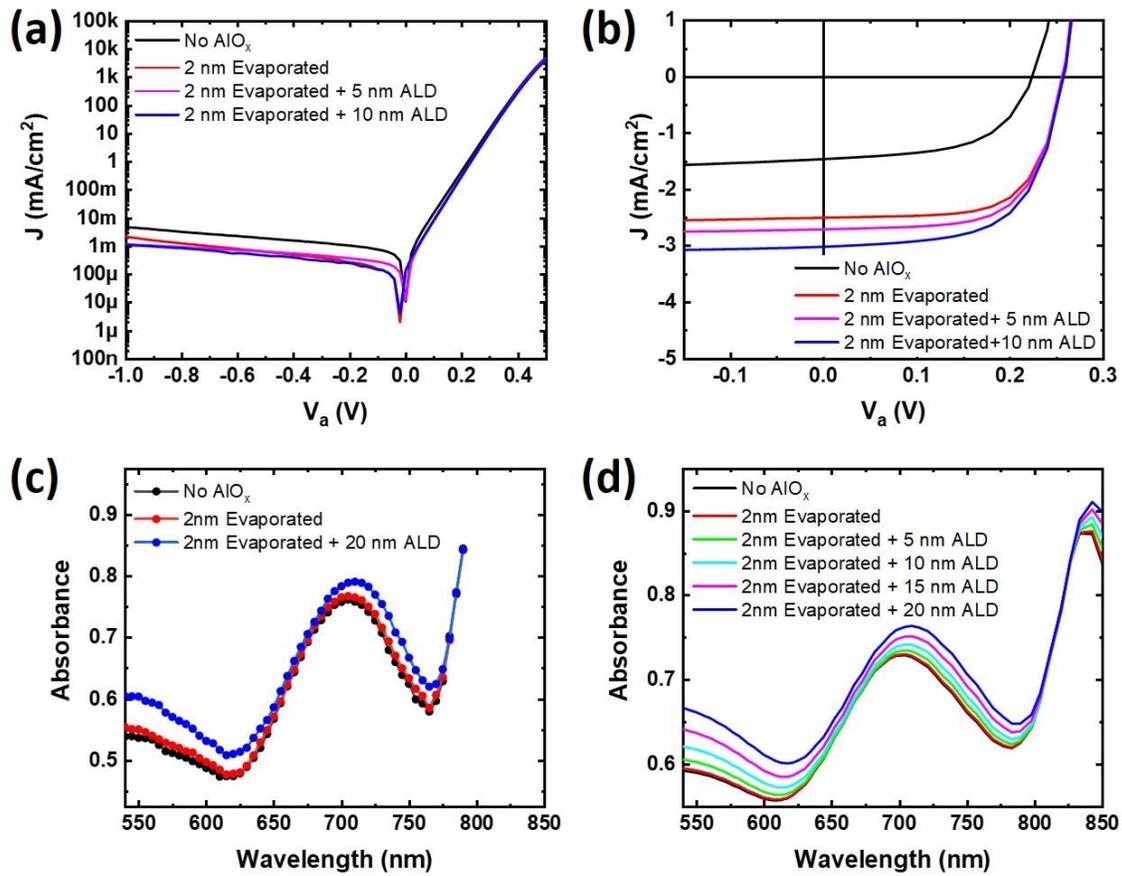

*Figure 2. (a) Dark current characteristics of the pristine and Al$_2$O$_3$ passivated devices for various thicknesses of Al$_2$O$_3$ deposited onto the 120 nm thick WSe$_2$ layer of device B. (b) Photovoltaic characteristics of the pristine and passivated device measured under a 30 mW/cm$^2$ 3400 K black body source. (c) Absorbance of the pristine and passivated device. (d) Simulated absorbance of the pristine and passivated device.*

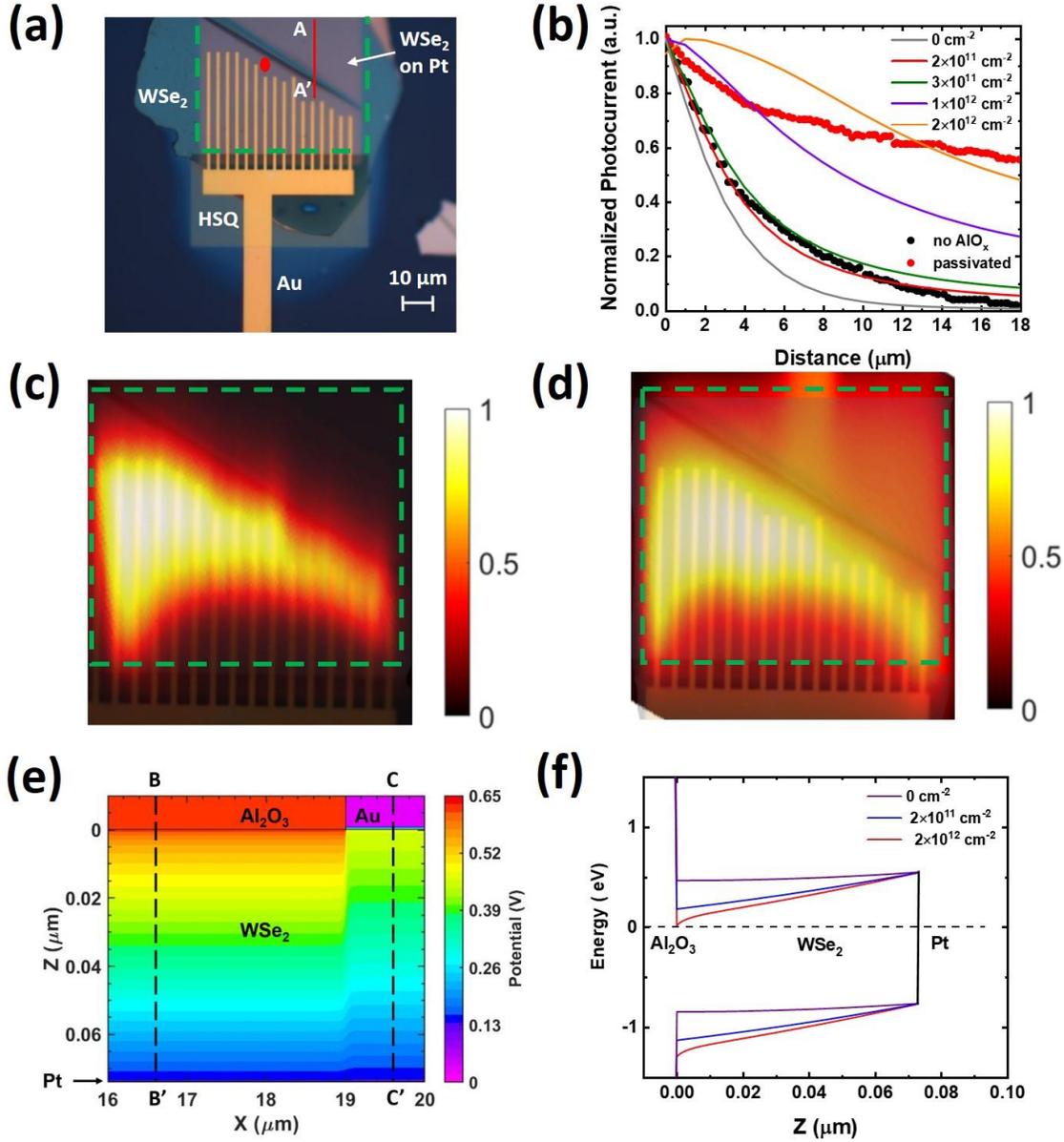

Figure 3. (a) Optical image of device C depicting point of measurement for the EQE and power dependence (red dot). The green dotted line outlines the active device area. The top electrode has a trapezoidal shape to avoid a wrinkle in the flake. (b) Normalized photocurrent (symbols) profile along A'A of the device in the pristine state and with $Al_2O_3$ passivation compared to simulated device characteristics (solid lines) (c) and (d) show the photocurrent map of pristine and passivated device, respectively, with a 1 μm spot size laser beam at 532 nm and 12 μW. The green dotted line shows the bottom platinum contact. (e) Simulated potential profile of the passivated device with $10^{12}$ cm$^{-2}$ surface charge. (f) Energy band diagram across BB' showing comparison between simulated devices as a function of positive surface charge density.

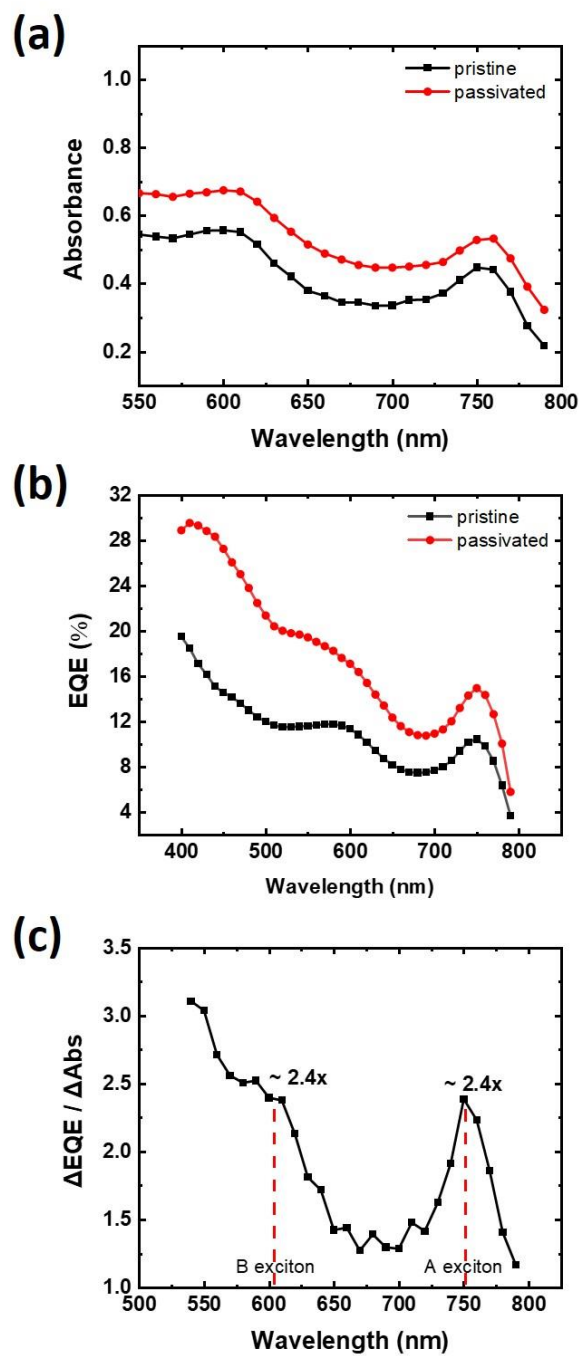

*Figure 4. (a) Absorption of the pristine and 20 nm ALD passivated 74 nm flake of device C as a function of wavelength. (b) EQE of the pristine and passivated device as a function of wavelength. Measurements were taken using a 1 μm beam spot size with approximately 20 μW power at the same location. (c) Comparison of the change in EQE to the change in absorption to highlight regions where the change in EQE is far greater than the change in absorption.*

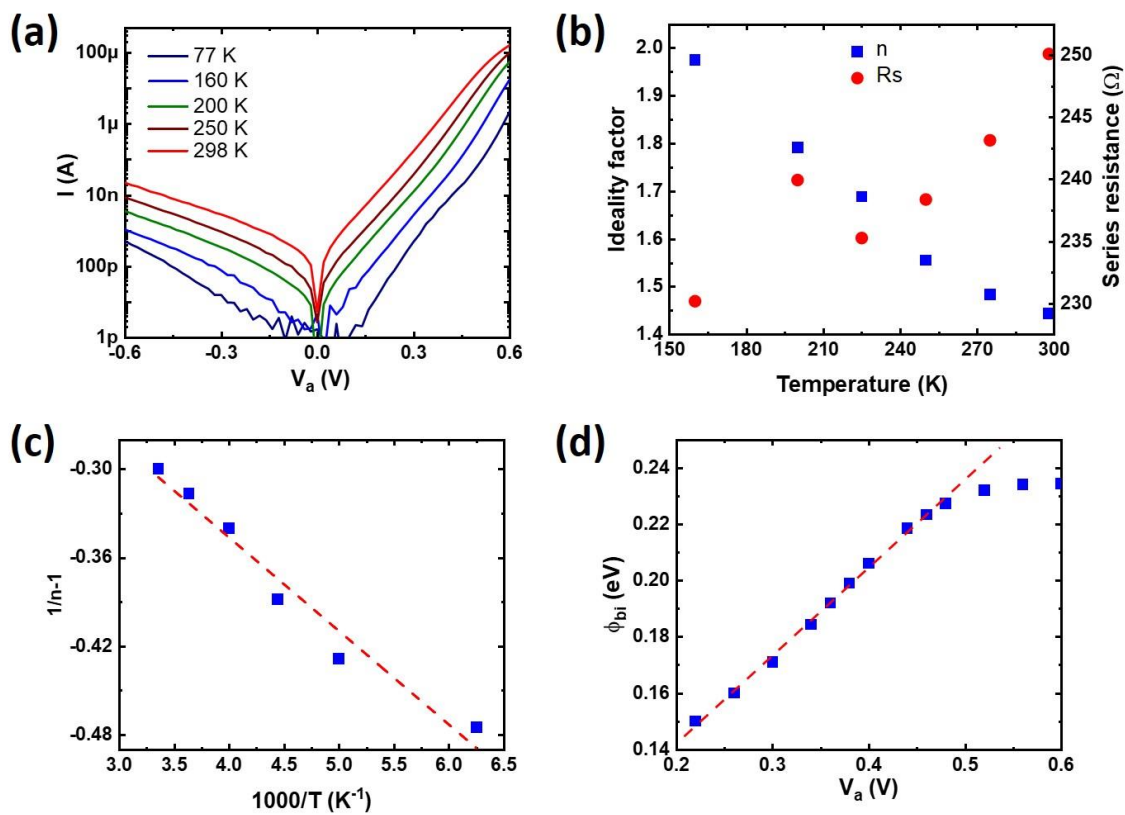

Figure 5. (a) Logarithmic plot of the device dark current vs. applied bias for pristine device D with a 25 nm WSe$_2$ flake. (b) Ideality factor and series resistance as a function of temperature. (c) Ideality factor analysis according to Werner's methods. (d) Extracted Schottky barrier heights as a function of the applied bias voltage.

AUTHOR INFORMATION

**Corresponding Author**

* Email: emcvay@mit.edu

**Author Contributions**

The manuscript was written through contributions of all authors. All authors have given approval to the final version of the manuscript. ‡These authors contributed equally.

ACKNOWLEDGMENT

The authors would like to thank Prof. Tonio Buonassisi and Prof. Pablo-Jarillo Herrero for providing research facilities. This work has been partially supported by the NASA NSTRF program, the STC Center for Integrated Quantum Materials, NSF Grant No. DMR-1231319, and the AFOSR FATE MURI, Grant No. FA9550-15-1-0514.

REFERENCES

1. Koman, V. B.; Liu, P.; Kozawa, D.; Liu, A. T.; Cottrill, A. L.; Son, Y.; Lebron, J. A.; Strano, M. S., Colloidal nanoelectronic state machines based on 2D materials for aerosolizable electronics. *Nat Nanotechnol* **2018**, *13* (9), 819-827.
2. Jariwala, D.; Davoyan, A. R.; Wong, J.; Atwater, H. A., Van der Waals Materials for Atomically-Thin Photovoltaics: Promise and Outlook. *Acs Photonics* **2017**, *4* (12), 2962-2970.
3. Jariwala, D.; Davoyan, A. R.; Tagliabue, G.; Sherrott, M. C.; Wong, J.; Atwater, H. A., Near-Unity Absorption in van der Waals Semiconductors for Ultrathin Optoelectronics. *Nano Lett* **2016**, *16* (9), 5482-7.
4. Wong, J.; Jariwala, D.; Tagliabue, G.; Tat, K.; Davoyan, A. R.; Sherrott, M. C.; Atwater, H. A., High Photovoltaic Quantum Efficiency in Ultrathin van der Waals Heterostructures. *ACS Nano* **2017**, *11* (7), 7230-7240.
5. Lin, P.; Zhu, L.; Li, D.; Xu, L.; Pan, C.; Wang, Z., Piezo-Phototronic Effect for Enhanced Flexible MoS2/WSe2 van der Waals Photodiodes. *Advanced Functional Materials* **2018**, 1802849.
6. Furchi, M. M.; Pospischil, A.; Libisch, F.; Burgdorfer, J.; Mueller, T., Photovoltaic effect in an electrically tunable van der Waals heterojunction. *Nano Lett* **2014**, *14* (8), 4785-91.
7. Kim, K.; Larentis, S.; Fallahazad, B.; Lee, K.; Xue, J. M.; Dillen, D. C.; Corbet, C. M.; Tutuc, E., Band Alignment in WSe2-Graphene Heterostructures. *Acs Nano* **2015**, *9* (4), 4527-4532.
8. Shockley, W.; Queisser, H. J., Detailed Balance Limit of Efficiency of p-n Junction Solar Cells. *Journal of Applied Physics* **1961**, *32* (3), 510-519.
9. Schmidt, H.; Giustiniano, F.; Eda, G., Electronic transport properties of transition metal dichalcogenide field-effect devices: surface and interface effects. *Chem Soc Rev* **2015**, *44* (21), 7715-36.

10. Frindt, R. F., The Optical Properties of Single Crystals of WSe2 and MoTe2. *J. Phys. Chem. Solids* **1963,** *24*, 1107-1112.
11. Choi, M. S.; Qu, D.; Lee, D.; Liu, X.; Watanabe, K.; Taniguchi, T.; Yoo, W. J., Lateral MoS2 p-n Junction Formed by Chemical Doping for Use in High-Performance Optoelectronics. *Acs Nano* **2014,** *8* (9), 9332-9340.
12. Tang, Y.; Wang, Z.; Wang, P.; Wu, F.; Wang, Y.; Chen, Y.; Wang, H.; Peng, M.; Shan, C.; Zhu, Z.; Qin, S.; Hu, W., WSe2 Photovoltaic Device Based on Intramolecular p-n Junction. *Small* **2019,** *15* (12), e1805545.
13. Wi, S. J.; Chen, M. K.; Li, D.; Nam, H.; Meyhofer, E.; Liang, X. G., Photovoltaic response in pristine WSe2 layers modulated by metal-induced surface-charge-transfer doping. *Applied Physics Letters* **2015,** *107* (6), 062102.
14. Saint-Cast, P.; Benick, J.; Kania, D.; Weiss, L.; Hofmann, M.; Rentsch, J.; Preu, R.; Glunz, S. W., High-Efficiency c-Si Solar Cells Passivated With ALD and PECVD Aluminum Oxide. *Ieee Electron Device Letters* **2010,** *31* (7), 695-697.
15. Dingemans, G.; Seguin, R.; Engelhart, P.; Sanden, M. C. M. v. d.; Kessels, W. M. M., Silicon surface passivation by ultrathin Al2O3films synthesized by thermal and plasma atomic layer deposition. *physica status solidi (RRL) - Rapid Research Letters* **2010,** *4* (1-2), 10-12.
16. Yu, L. L.; Zubair, A.; Santos, E. J. G.; Zhang, X.; Lin, Y. X.; Zhang, Y. H.; Palacios, T., High-Performance WSe2 Complementary Metal Oxide Semiconductor Technology and Integrated Circuits. *Nano Letters* **2015,** *15* (8), 4928-4934.
17. Pettersson, L. A. A.; Roman, L. S.; Inganas, O., Modeling photocurrent action spectra of photovoltaic devices based on organic thin films. *Journal of Applied Physics* **1999,** *86* (1), 487-496.
18. Li, Y.; Chernikov, A.; Zhang, X.; Rigosi, A.; Hill, H. M.; van der Zande, A. M.; Chenet, D. A.; Shih, E.-M.; Hone, J.; Heinz, T. F., Measurement of the optical dielectric function of monolayer transition-metal dichalcogenides:MoS2,MoSe2,WS2, andWSe2. *Physical Review B* **2014,** *90* (20).
19. Islam, R.; Saraswat, K., Limitation of Optical Enhancement in Ultra-thin Solar Cells Imposed by Contact Selectivity. *Sci Rep* **2018,** *8* (1), 8863.
20. Kufer, D.; Konstantatos, G., Highly Sensitive, Encapsulated MoS2 Photodetector with Gate Controllable Gain and Speed. *Nano Lett* **2015,** *15* (11), 7307-13.
21. Rehman, A. U.; Khan, M. F.; Shehzad, M. A.; Hussain, S.; Bhopal, M. F.; Lee, S. H.; Eom, J.; Seo, Y.; Jung, J.; Lee, S. H., n-MoS2/p-Si Solar Cells with Al2O3 Passivation for Enhanced Photogeneration. *ACS Appl Mater Interfaces* **2016,** *8* (43), 29383-29390.
22. Yang, S.; Cha, J.; Kim, J. C.; Lee, D.; Huh, W.; Kim, Y.; Lee, S. W.; Park, H. G.; Jeong, H. Y.; Hong, S.; Lee, G. H.; Lee, C. H., Monolithic Interface Contact Engineering to Boost Optoelectronic Performances of 2D Semiconductor Photovoltaic Heterojunctions. *Nano Lett* **2020**.
23. Simon, D. K.; Jordan, P. M.; Mikolajick, T.; Dirnstorfer, I., On the Control of the Fixed Charge Densities in Al2O3-Based Silicon Surface Passivation Schemes. *ACS Appl Mater Interfaces* **2015,** *7* (51), 28215-22.
24. Graham, R.; Miller, C.; Oh, E.; Yu, D., Electric field dependent photocurrent decay length in single lead sulfide nanowire field effect transistors. *Nano Lett* **2011,** *11* (2), 717-22.
25. Xiao, R.; Hou, Y.; Fu, Y.; Peng, X.; Wang, Q.; Gonzalez, E.; Jin, S.; Yu, D., Photocurrent Mapping in Single-Crystal Methylammonium Lead Iodide Perovskite Nanostructures. *Nano Lett* **2016,** *16* (12), 7710-7717.

26. Qiannan Cui, F. C., Nardeep Kumar, and Hui Zhao, Transient Absorption Microscopy of Monolayer and Bulk WSe2. *ACSNano* **2014,** *8* (3), 2970-2976.
27. Chen, X.; Wang, Z.; Wang, L.; Wang, H. Y.; Yue, Y. Y.; Wang, H.; Wang, X. P.; Wee, A. T. S.; Qiu, C. W.; Sun, H. B., Investigating the dynamics of excitons in monolayer WSe2 before and after organic super acid treatment. *Nanoscale* **2018,** *10* (19), 9346-9352.
28. Movva, H. C. P.; Rai, A.; Kang, S.; Kim, K.; Fallahazad, B.; Taniguchi, T.; Watanabe, K.; Tutuc, E.; Banerjee, S. K., High-Mobility Holes in Dual-Gated WSe2 Field-Effect Transistors. *Acs Nano* **2015,** *9* (10), 10402-10410.
29. Miller, E. J.; Yu, E. T.; Waltereit, P.; Speck, J. S., Analysis of reverse-bias leakage current mechanisms in GaN grown by molecular-beam epitaxy. *Applied Physics Letters* **2004,** *84* (4), 535-537.
30. Wittmer, M., Conduction mechanism in PtSi/Si Schottky diodes. *Phys Rev B Condens Matter* **1991,** *43* (5), 4385-4395.
31. Güttler, J. H. W. H. H., Transport Properties of Inhomogeneous Schottky

Contacts. *Physica Scripta* **1991,** *T39*, 258-264.
32. Cora M. Went, J. W., Phillip R. Jahelka, Michael Kelzenberg, Souvik Biswas, Harry; Atwater, A., A new metal transfer process for van der Waals contacts to vertical Schottky-junction transition metal dichalcogenide photovoltaics. *arXiv:1903.08191* **2019**.
33. Laurent, M. A.; Gupta, G.; Suntrup, D. J.; DenBaars, S. P.; Mishra, U. K., Barrier height inhomogeneity and its impact on (Al,In,Ga)N Schottky diodes. *Journal of Applied Physics* **2016,** *119* (6), 064501.
34. Lee, J. I.; Brini, J.; Dimitriadis, C. A., Simple parameter extraction method for non-ideal Schottky barrier diodes. *Electron Lett* **1998,** *34* (12), 1268-1269.
35. Nourbakhsh, A.; Zubair, A.; Dresselhaus, M. S.; Palacios, T., Transport Properties of a MoS2/WSe2 Heterojunction Transistor and Its Potential for Application. *Nano Lett* **2016,** *16* (2), 1359-66.
36. Zhu, Y.; Zhou, R.; Zhang, F.; Appenzeller, J., Vertical charge transport through transition metal dichalcogenides - a quantitative analysis. *Nanoscale* **2017,** *9* (48), 19108-19113.